\documentclass[aps,groupedaddress,nofootinbib,superscriptaddress,twopage,letterpaper]{revtex4}

\usepackage{epsfig}
\usepackage{multirow}

\begin{document}

\title{Scaling study of dynamical smeared-link clover fermions}

\preprint{CPT-P012-2008}

\author{S.~D\"urr}
\affiliation{NIC, DESY Zeuthen, D-15738 Zeuthen and FZ J\"ulich,D-52425 J\"ulich, Germany}
\author{Z.~Fodor}
\affiliation{NIC, DESY Zeuthen, D-15738 Zeuthen and FZ J\"ulich,D-52425 J\"ulich, Germany}
\affiliation{Bergische Universit\"at Wuppertal, Gaussstr.\,20, D-42119 Wuppertal, Germany}
\affiliation{Institute for Theoretical Physics, E\"otv\"os University, H-1117 Budapest, Hungary}
\author{C.~Hoelbling}
\author{R.~Hoffmann}
\affiliation{Bergische Universit\"at Wuppertal, Gaussstr.\,20, D-42119 Wuppertal, Germany}
\author{S.~D.~Katz}
\affiliation{Bergische Universit\"at Wuppertal, Gaussstr.\,20, D-42119 Wuppertal, Germany}
\affiliation{Institute for Theoretical Physics, E\"otv\"os University, H-1117 Budapest, Hungary}
\author{S.~Krieg}
\affiliation{J\"ulich Supercomputing Center, FZ J\"ulich, D-52425 J\"ulich, Germany}
\author{T.~Kurth}
\affiliation{Bergische Universit\"at Wuppertal, Gaussstr.\,20, D-42119
Wuppertal, Germany}
\author{L.~Lellouch}
\affiliation{Centre de Physique Th\'eorique\,\footnote{\scriptsize CPT is ``UMR 6207 du CNRS et des universit\'es d'Aix-Marseille I, d'Aix-Marseille II et du Sud Toulon-Var, affili\'ee \`a la FRUMAM''.}, Case 907, Campus de Luminy, F-13288 Marseille Cedex 9, France}
\author{T.~Lippert}
\affiliation{NIC, DESY Zeuthen, D-15738 Zeuthen and FZ J\"ulich,D-52425 J\"ulich, Germany}
\affiliation{Bergische Universit\"at Wuppertal, Gaussstr.\,20, D-42119 Wuppertal, Germany}
\affiliation{J\"ulich Supercomputing Center, FZ J\"ulich, D-52425 J\"ulich, Germany}
\author{K.K.~Szabo}
\affiliation{Bergische Universit\"at Wuppertal, Gaussstr.\,20, D-42119 Wuppertal, Germany}
\author{G.~Vulvert}
\affiliation{Centre de Physique Th\'eorique\,\footnote{\scriptsize CPT is ``UMR 6207 du CNRS et des universit\'es d'Aix-Marseille I,
d'Aix-Marseille II et du Sud Toulon-Var, affili\'ee \`a la FRUMAM''.}, Case 907, Campus de Luminy,
        F-13288 Marseille Cedex 9, France}

\collaboration{Budapest-Marseille-Wuppertal Collaboration}

\date{February 19, 2008}

\begin{abstract}
We present a framework for phenomenological lattice QCD calculations
which makes use of a tree level Symanzink improved action for gluons
and stout-link Wilson fermions. We give details of our efficient
HMC/RHMC algorithm and present a scaling study of the low-lying
$N_f=3$ baryon spectrum. We find a scaling region that extends to
$a\lesssim 0.16$~fm and conclude that
our action and algorithm are suitable for large scale
phenomenological investigations of $N_f=2+1$ QCD. We expect this
conclusion to hold for other comparable actions.
\end{abstract}


\maketitle


\hyphenation{conti-nu-um}


\section{Introduction}


Over the last decade, it has become clear that smeared-link fermion
actions\footnote{In the literature they are also referred to as
  ``UV-filtered'' or ``fat-link'' actions. Likewise, actions in which
  the covariant derivative involves the original gauge links are
  sometimes called ``thin-link'' actions.}  offer substantial
technical advantages over their thin-link counterparts.  The idea of
damping unphysical UV fluctuations by replacing elementary links with
a weighted sum of paths was first introduced in the framework of pure
gauge theory~\cite{Albanese:1987ds}.  It was later recognized that the
chiral properties of clover fermions~\cite{Sheikholeslami:1985ij} can
be substantially improved by replacing the thin links in the covariant
derivative of the fermion operator with their smeared
counterparts~\cite{DeGrand:1998jq}. From a Symanzik point of view,
this replacement amounts to adding ultralocal irrelevant terms to the
fermion action, as long as the smearing prescription (parameter,
iteration number) stays fixed as a function of bare coupling. In this
way it is guaranteed that the continuum limit is unchanged.

In the context of quenched QCD, the advantages of smeared clover
fermions are well
established~\cite{DeGrand:1998jq,Bernard:1999kc,Stephenson:1999ns,Bernard:2000ht,Zanotti:2001yb,Capitani:2006ni}. The
theoretically leading $O(\alpha_s a)$ contributions are, in practice,
absent and the extrapolation to the continuum appears to be dominated
by $O(a^2)$ cut-off effects. In particular, the tamed UV fluctuations
result in improved chiral symmetry properties. Furthermore, the
smearing significantly reduces the contributions of unphysical
tadpoles; renormalization constants are generally closer to their tree
level values, and $c_\mathrm{SW}$ is not far from $1$ at typical
lattice spacings.

Given this experience, it is reasonable to expect that also dynamical
clover fermions will benefit from link smearing. There, the
non-differentiable nature of the back projection step of the smeared
link onto the gauge group, which is usually performed, for instance,
when using APE smearing~\cite{Albanese:1987ds}, may pose problems for
the molecular dynamics update.  An early suggestion was to use
``stout'' links~\cite{Morningstar:2003gk} to define fermions which can
be simulated with the Hybrid Monte Carlo (HMC)
algorithm~\cite{Duane:1987de}. Further particulars of the HMC force
with UV-filtered actions have been worked out in~\cite{Kamleh:2004xk}.
Recently, several alternative smearing methods suitable for dynamical
simulations have been
proposed~\cite{Capitani:2006ni,Hasenfratz:2007rf,Durr:2007cy,Schaefer:2007dc,Hoffmann:2007nm,Moran:2008ra}.

The efficiency of link smearing results from the fact that it leaves
the structure of the fermionic operator entirely unchanged. Smeared
clover fermions still have exclusively nearest-neighbour
couplings. The damping of unphysical UV modes is achieved exclusively
by a modified - but still ultralocal - coupling to the gluonic
background. This modification of the fermionic action is continuum
irrelevant, but at a given finite cutoff one generally expects
observables with weaker coupling to unphysical UV modes to be closer
to their continuum limit values, resulting in overall improved
scaling. In the present paper we investigate this issue by performing
a scaling study with $N_f=3$, stout-link clover fermions. Although
other smearing methods are presently known, we opt for the standard
stout-link prescription because it is widely used and will share
features, such as an enlarged scaling region, with other comparable
prescriptions.

The size of its scaling region is one of the most important criteria
to assess the suitability of a given action for phenomenological
purposes. The onset of scaling, together with the power of the lattice
spacing against which results need to be plotted to show a linear
dependence, determines the finest lattice spacing needed to reliably
extrapolate to the continuum, and hence the overall cost in terms of
CPU time. Our main result is that smeared clover fermions do indeed
show very nice scaling properties up to at least $0.16$~fm lattice
spacing. Moreover, the link-smearing seems to eliminate known
pathologies that unfiltered actions may show in a dynamical
setting~\cite{DellaMorte:2004hs,Aoki:2004iq}.

The results presented in this paper are obtained using a tree-level
Symanzik improved gauge action~\cite{Luscher:1985zq} and six-step,
stout-smeared clover fermions with a clover coefficient taken at its
tree-level value $c_\mathrm{SW}=1$ (though a
perturbative~\cite{Wohlert:1987rf,Horsley:2007fw} or
non-perturbative~\cite{Luscher:1996ug,Hoffmann:2007nm} determination is
feasible). Note that also $F_{\mu\nu}$ in the clover term is built
from the same set of stout links.
This choice allows for efficient simulation while delivering good
scaling properties, as demonstrated below. Moreover, dedicated studies
in quenched QCD have shown that the dependence of observables on
smearing is quite mild (see e.g.\ \cite{Capitani:2006ni,Durr:2007cy}) and the exploratory studies of
e.g.\ \cite{Kamleh:2004xk,Hasenfratz:2007rf,Schaefer:2007dc,Hoffmann:2007nm,Moran:2008ra} 
suggest that this behavior persists in the the full
theory. Thus, our choice involves no fine-tuning and we expect our
results to hold for actions which involve comparable amounts of smearing.

In our scaling study we choose $N_f=3$ for simplicity,
creating an artificial world with degenerate $u$, $d$ and $s$
quarks. We will denote the pseudoscalar and vector mesons by $\pi$ and
$\rho$ respectively. Our goal is to perform continuum extrapolations
along three distinct lines of constant ``physical'' quark masses,
characterized by $M_\pi/M_\rho=0.60,0.64\text{ and }0.68$.  Since we
do not aim in this paper at phenomenologically relevant computations
and instead would like to test the extent of the scaling regime, we
deliberately choose these rather large masses. With $Ma$ for standard
hadrons close to one, cut-off effects with inferior actions will be
large. In all our runs $M_\pi L$ is kept fixed, at values larger than
four, to avoid finite volume effects. Our goal is to simulate at
several values of the gauge coupling and fixed $M_\pi/M_\rho$ and
$M_\pi L$, and to determine the scaling of the baryon octet and
decuplet masses, $M_{\rm N}$ and $M_\Delta$.

The remainder of the article is organized as follows.  In Sec.\,2
details of the action and our algorithm are given.  Secs.\,3 and 4 are
devoted to tests which provide clear evidence for the absence of bulk
phase transitions in our simulations. In Sec.\,5 we show that our action is 
ergodic with respect to topology.  Sec.\,6 then contains a detailed
scaling study of the nucleon and delta masses.  We conclude with a
short summary and outlook.


\section{Action and algorithms}


\begin{figure}
\epsfig{file=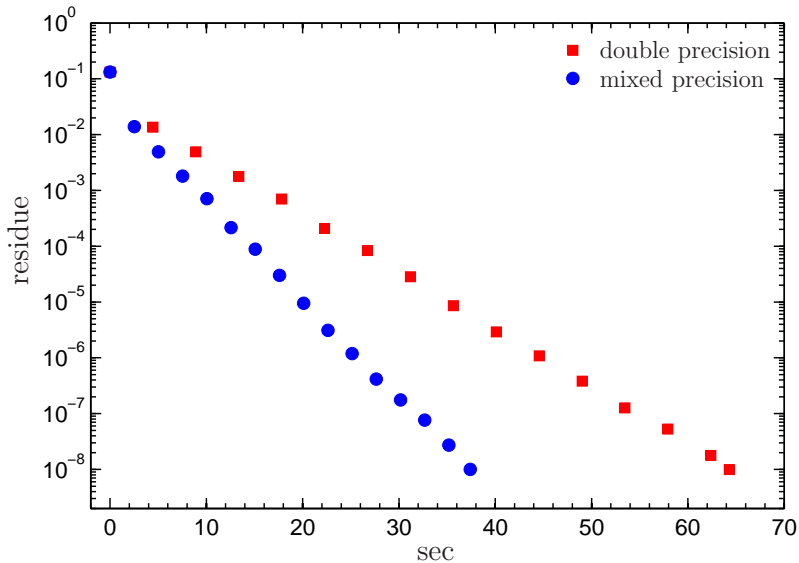,scale=0.78}
\caption{Performance of CG in double precision (squares) compared
  to a mixed precision variant of CG (circles). Data are from an
  $N_f=2+1$ run on a $32^3\times 64$ lattice at $\beta=3.57$ with
  $am_{ud}^\mathrm{PCAC}\simeq 0.0077$ and $am_{s}^\mathrm{PCAC}\simeq
  0.049$ corresponding to $M_\pi\sim 250$~MeV.
\label{fig:cg}}
\end{figure}

\subsection{Action}
The explicit form of our gauge and fermion action in terms of the thin
($U_{n,\mu}$) and smeared ($V_{n,\mu}$) gauge links is as follows:
\begin{eqnarray}
S  &=& S_G^\mathrm{Sym} + S_F^{\mathrm{SW}} \nonumber\\
S_G^\mathrm{Sym} & = & \beta\,\,\Big[\frac{c_0}{3}\,
\sum_\mathrm{plaq} {\rm Re\, Tr\,}(1-U_\mathrm{plaq})
\, + \frac{c_1}{3}\, \sum_\mathrm{rect} {\rm Re \, Tr\,}
(1- U_\mathrm{rect})\Big]\\
S_F^{\mathrm{SW}} &=& S_F^\mathrm{W}[V]- \frac{c_{\mathrm{SW}}}{4}\,\sum_n\,\sum_{\mu,\nu}
\overline{\psi}_x\,\sigma_{\mu\nu}F_{\mu\nu,n}[V]\,\psi_x\,, \nonumber
\end{eqnarray}
with the standard Wilson action $S_F^\mathrm{W}$. The parameters $c_{\mathrm{SW}}, c_0$ and $c_1$ set to their tree level values: 
$$c_{\mathrm{SW}} = 1, \quad c_1=-1/12, \quad
c_0=1-8c_1=5/3\,.$$ 
Both the hopping part and the clover improvement term in the fermion action $S_F^{\mathrm{SW}}$
use six-step stout-smeared links~\cite{Morningstar:2003gk} $V_{n,\mu}\equiv V^{(6)}_{n,\mu}$.
Those are constructed from the thin links $U_{n,\mu}\equiv V^{(0)}_{n,\mu}$ according to
\begin{eqnarray}\label{eq:smearing}
V^{(n+1)}& = & e^{\rho S^{(n)}}U^{(n)},\label{eq:def-stout}\nonumber\\
S^{(n)} & = & \frac{1}{2}(\Gamma^{(n)} V^{(n)\dagger}-V^{(n)}\Gamma^{(n)\dagger})-\frac{1}{6}{\rm Re \,Tr}(\Gamma^{(n)}
V^{(n)\dagger}-V^{(n)}\Gamma^{(n)\dagger})\\
\Gamma^{(n)}_{n,\mu}&=&\sum_{\nu\neq\mu}V^{(n)}_{n,\nu}V^{(n)}_{n+\nu,\mu}V^{(n)\dagger}_{n+\mu,\nu}\nonumber
\end{eqnarray}
The stout smearing parameter is chosen to be $\rho=0.11$, which is a
rather conservative choice~\cite{Morningstar:2003gk,Capitani:2006ni} corresponding to
an $\alpha_\mathrm{APE}=0.48$ with respect to the average
plaquette~\cite{Hasenfratz:2007rf}. In $S_G^\mathrm{Sym}$ only the
unfiltered links are used. As detailed in the Introduction, this
action is ultralocal in both the quark and gauge sector.

\subsection{Simulation algorithm}

We start with the description of our $N_f=2+1$ algorithm. Two flavors
are implemented via the Hybrid Monte Carlo (HMC)
algorithm~\cite{Duane:1987de}, the third using the Rational Hybrid
Monte Carlo (RHMC) algorithm~\cite{Clark:2002vz,Clark:2006fx}. We employ even/odd
preconditioning~\cite{DeGrand:1990dk} to speed up the fermion matrix
inversions. The generic HMC algorithm suffers from critical slowing
down in the light-quark regime. To treat this problem, we combine
several improvements over the generic algorithm (see
also~\cite{AliKhan:2003br,Urbach:2005ji}):

\begin{itemize}
	
\item {\it Multiple time-scale integration:} not all force
  contributions in the molecular dynamics (MD) part of the HMC
  algorithm require the same amount of computational resources. Using
  multiple time-scale integration (``Sexton-Weingarten integration
  scheme'')~\cite{Sexton:1992nu}, it is possible to put each part of
  the MD on a different time scale according to its relative
  contribution to the total force, thus reducing the computational
  costs of the MD.
 
\item {\it Mass preconditioning:} the pseudofermion force is used
  within the MD to include the effects of dynamical fermions. Through
  mass preconditioning, the UV part of the force can be split off and
  treated separately~\cite{Hasenbusch:2001ne}, which helps reducing the
  fluctuations in the force. The second important benefit of mass
  preconditioning appears when combined with the multiple timescale
  integration scheme~\cite{AliKhan:2003br,Urbach:2005ji}: the more
  expensive infrared part contributes less to the total force and can
  be integrated with larger time steps.

\item {\it RHMC:} the third, unpaired quark flavor is implemented
  through the RHMC~\cite{Clark:2002vz,Clark:2006fx} algorithm. This
  algorithm makes use of the fact that the single fermion action can
  be written as $\xi^\dag(M^\dag M)^{-1/2}\xi$, where the inverse
  square root can in turn be approximated by a rational approximation
  and be efficiently calculated with a multi-shift solver. The RHMC is
  highly efficient in simulating a single quark flavor. It can also be
  combined with the multiple timescale integration scheme.

\item {\it Omelyan integrator:} the MD integration within the generic
  HMC algorithms uses the leapfrog integration scheme. It proceeds by
  first integrating one half step in position space followed by a full
  step update of the conjugate momenta and finally another half step
  in position space. The Omelyan integrator adds a small momentum
  update (reduced by $\lambda\approx0.193$) before and after the
  leapfrog step and shortens the original leapfrog momentum update in
  by a factor $(1-2\lambda)$. This scheme improves the MD energy
  conservation by about one order of magnitude for a factor $\sim2$
  increase in computational cost. The use of a correspondingly
  larger step size then results in a net gain of about
  50\%~\cite{Takaishi:2005tz}.

\end{itemize}

We use this algorithm also for our $N_f=3$ scaling study with
$m_\mathrm{HMC}=m_\mathrm{RHMC}$.

\subsection{Inversion algorithms}

The most time consuming part, both in the valence and the sea sector,
is the (approximate) fermion matrix inversion by means of a linear
solver.  These calculations generally require double precision
accuracy.  This is due to the fact that, in order to maintain
reversibility, the MD part of the algorithm has to be performed in
double precision.  Double precision accuracy is also required in
valence calculations at small quark masses, owing to the large
condition numbers involved.  However, this does not imply that each
fermion matrix multiplication needs to be done in double precision.
In the valence sector we need to solve
\begin{equation}
Dx = b
\label{eqn:propagator}
\end{equation}
(with $D$ in our case being the stout-link clover Dirac operator) to construct the correlators.
To calculate the fermionic force in the MD part of the algorithm we need to
solve
\begin{equation}
D^\dagger Dx = b.
\label{eqn:determinant}
\end{equation}
In both cases it is possible to use a single precision version of $D$
within mixed precision solvers to accelerate the inversion. There is basically
no penalty in terms of the iteration count: we find that the increase
in the number of matrix multiplications is well below 10\%.

A simple and reasonably efficient way to construct a mixed precision
solver is to use the standard ``iterative refinement'' technique,
which amounts to repeatedly using a single precision solver. In this
scheme, only the (outer) residuals and global sums are calculated in
double precision; the inversion is performed with single precision
accuracy.  The single precision inversion typically uses the same
algorithm that would be used for a full double precision inversion,
such as BiCGstab to solve~(\ref{eqn:propagator}) or CG
for~(\ref{eqn:determinant}).  With $\mathbf{A}=D$ or
$\mathbf{A}=D^\dagger D$ referring to the forward multiplication
routine in double precision, $\mathbf{a}$ the single precision
counterpart and $\epsilon$ the desired final double precision
accuracy, the complete procedure reads:
\begin{enumerate}
\itemsep-2pt
\item Compute $r_i=b-\mathbf{A}x_i$
\item If $|r_i|\le\epsilon|b|$, exit
\item
Solve $\mathbf{a}t_i=r_i$ in single precision to an accuracy
$\epsilon^\prime$, with $\tilde{t}_i$ denoting the solution.
\item Update $x_{i+1}=x_i+\tilde{t}_i$
\item Goto 1
\end{enumerate}
With $s_i=r_i-\mathbf{A}\tilde{t}_i$ and
$\delta\equiv|s_i|/|r_i|\approx\epsilon'<1$, we have
\begin{equation}
|r_{i+1}|=|b-\mathbf{A}x_{i+1}|=|b-\mathbf{A}x_i-\mathbf{A}\tilde{t}_i|=
|b-\mathbf{A}x_i-r_i+s_i|=|s_i|=\delta|r_i|<|r_i|
\;.
\end{equation}
Thus, as long as the single precision inversion does not fail, the
method will converge. Since many single precision matrix
multiplications are needed to compute $\tilde{t}_i$, compared to just
one double precision multiplication with $\mathbf{A}$ in the outer
iteration, the whole solver is dominated by the single precision
matrix multiplication performance, resulting in a significant speedup
over a full double precision inversion (see Fig.\ \ref{fig:cg}).

\begin{figure}
\center
\epsfig{file=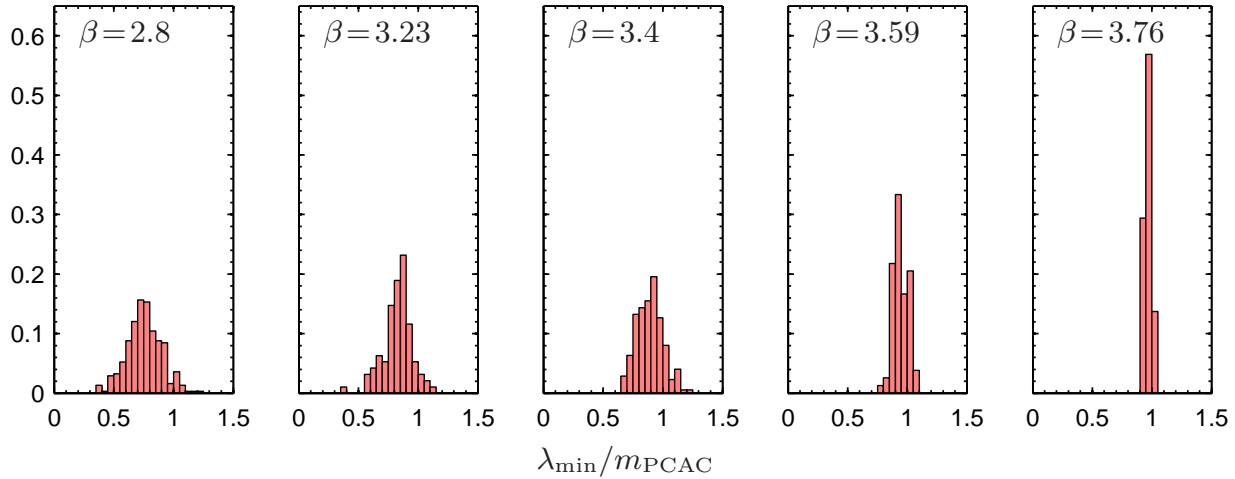,scale=1.3}
\vspace*{-2mm}
\caption{The magnitude of the smallest eigenvalue of the preconditioned
hermitean Dirac operator in units of the PCAC mass. At each $\beta$ the lightest
run ($M_\pi/M_\rho\simeq0.6$) is shown. \label{fig:spectralgap}}
\end{figure}


\section{Spectral gap}


In quenched QCD, the (unsmeared) clover fermion operator may have one
or several eigenvalues close to the origin or with a negative real
part, even for not very light quark masses. Configurations for which
this is the case are referred to as ``exceptional''.

If one integrated the HMC trajectories exactly, any such configuration
would be absent in full QCD, since an eigenvalue of the hermitean
Wilson operator $H_\mathrm{W}=\gamma_5 D_\mathrm{W}$ approaching zero
would induce an infinite back-driving force in the HMC.  In practice,
when the trajectories are generated with a finite step-size integrator,
the near zero modes along a trajectory are only approximately
suppressed. This may cause a breakdown of the MD evolution. It is
therefore natural to monitor the smallest eigenvalue (in magnitude) of
$H_\mathrm{W}$ and check if it is sufficiently far from the origin
throughout the entire run. In a given ensemble this spectral gap shows
a more-or-less Gaussian distribution, and as long as its median is
several $\sigma$ away from zero, the simulation is deemed
safe~\cite{DelDebbio:2005qa}.

Since we use even-odd preconditioning, the relevant quantity to
monitor is the smallest eigenvalue of the hermitean counterpart of the
reduced operator
$D_\mathrm{red}={1\over2}(D_\mathrm{oo}-D_\mathrm{oe}^{}D_\mathrm{ee}^{-1}D_\mathrm{eo}^{})$,
which is $\gamma_5$-hermitean. We include a factor $1/2$ to have its
IR eigenvalues almost aligned with the low-lying eigenvalues of the
full operator. For the lightest mass ($M_\pi/M_\rho=0.60$,
cf.\ Sect.\,5) the distributions are shown in
Fig.\,\ref{fig:spectralgap}, with $\beta$ ranging from 2.8 (left) to
3.76 (right). One can see that even for the strongest coupling, there
is still a clear separation of the eigenmodes from the origin.

\begin{figure}
\center
\epsfig{file=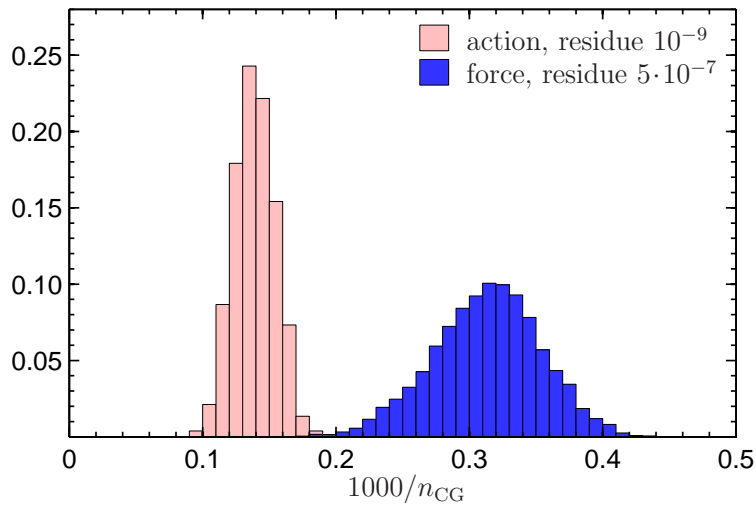,scale=0.9}
\vspace*{-2mm}
\caption{\label{fig:inviter}Histogram of the inverse iteration number of our linear solver
  at a lighter $M_\pi$ for the lightest pseudofermion in the action. Results are from an $N_f = 2+1$ run on a
  $48^3\times 64$ lattice at $\beta = 3.57$ with $am_{ud}^{PCAC}\simeq
  0.0056$ and $am_{s}^{PCAC}\simeq 0.044$ corresponding to $M_\pi\sim
  190$~MeV.
}
\end{figure}

For phenomenological applications it is of course most relevant to
know how this spectral gap evolves when lowering the masses of two of
the three flavors. Instead of monitoring the lowest eigenvalue of
$\gamma_5 D_\mathrm{red}$, we opted for monitoring the closely related
quantity $1/n_\mathrm{CG}$, where $n_\mathrm{CG}$ is the iteration
count for the lightest pseudofermion in the action for our $N_f=2+1$
runs. In Fig.~\ref{fig:inviter}, we plot a histogram of
$1/n_\mathrm{CG}$ for one of our lightest production runs (for
phenomenological studies) and find a clear gap, which provides strong
evidence for the stability of the algorithm. We have also monitored
the acceptance rate and the Hamiltonian violation $\Delta H$
throughout our runs and have seen no sign of any algorithmic problems.


\section{Search for potentially metastable behavior}


In dynamical Wilson fermion simulations with small quark masses, it
was reported that the system appears to undergo a first-order
transition to an unphysical
phase~\cite{Aoki:2004iq,Farchioni:2004us}. This was argued to mean
that there is a lower bound on the quark mass, below which physically
sensible simulations cannot be performed. Moreover, it was observed,
that
\begin{enumerate}
\item the phenomenon occurs only with coarse lattices,
\item gauge action improvement decreases the lower bound on the quark
mass~\cite{Farchioni:2004fs},
\item $O(a)$-improved Wilson fermions together with improved gauge
actions made the problem disappear for all lattice spacings investigated
in~\cite{Aoki:2004iq},
\item one level of stout smearing weakens the phenomenon~\cite{Jansen:2007sr}.
\end{enumerate}

When discussing such phenomena, it is important to remember that a
first-order phase transition can only occur in infinite volume. In
finite volume, the metastability can be understood as an artifact of
the updating algorithm: with an efficient algorithm, the system should
eventually find the true minimum of the effective potential. Thus, for
finite-volume simulations, the relevant question is: can the algorithm
thermalize the system in a manageable number of updating steps?

\begin{figure}
\begin{center}
\epsfig{file=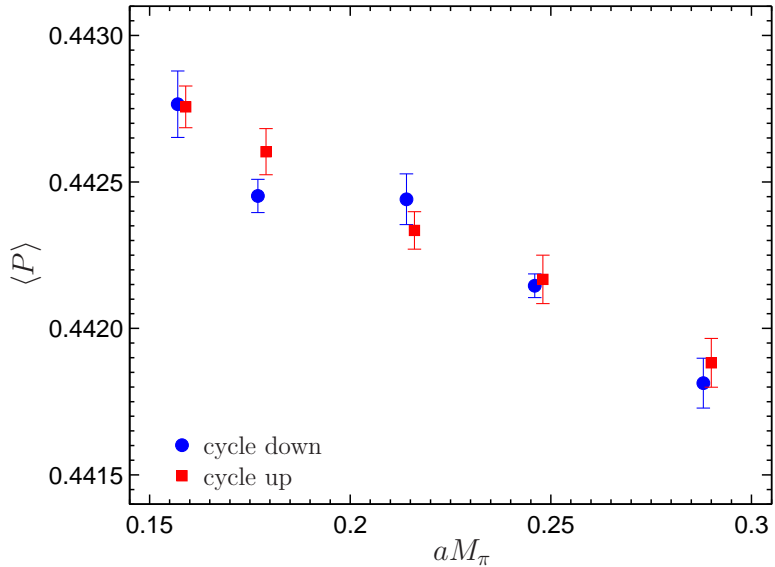,scale=0.8}
\caption{Absence of hysteresis in the average expectation value of the
  plaquette. Data are from an $N_f=2+1$ run on a $16^3\times 32$
  lattice at $\beta=3.3$ with a fixed strange quark mass
  $am_s^\mathrm{PCAC}\simeq 0.0677$ and the light quark mass varying
  between $am_{ud}^\mathrm{PCAC}\simeq0.0066$ and
  $0.0243$ in ascending (square) and
  descending (circles) order. The range of light quark masses
  corresponds to $M_\pi\sim 240-440$~MeV. The second data set is
  slightly offset along the $x$-axis for better readability.
\label{hyst}}
\end{center}
\end{figure}

To investigate this issue, we have taken two $16^3\times 32$
configurations, one with random links and the other, thermalized in a
$N_f=2+1$ simulation at $\beta=3.3$, with
$am^\mathrm{PCAC}_{u,d}=0.0066$, corresponding to a pion mass of
approximately 240~MeV, and $am^\mathrm{PCAC}_s\simeq 0.0677$,
corresponding roughly to the physical strange quark mass.  A
``downward'' updating sequence was then constructed from the random
configuration: consecutive simulations at
$am^\mathrm{PCAC}_{u,d}\simeq 0.0243, 0.0173, 0.0131, 0.0086, 0.0066$,
corresponding to a range of pseudoscalar masses $M_\pi\sim
440-240$~MeV, were performed, with each simulation starting from the
last configuration of the previous (larger mass) run.  Similarly, an
``upward'' sequence of five simulations was obtained, beginning with
the configuration thermalized at $am^\mathrm{PCAC}_{u,d}\simeq
0.0066$, and ending with a run at $am^\mathrm{PCAC}_{u,d}\simeq
0.0243$. For each point in the two sequences, approximately 400
trajectories were generated, of which the first 100 were discarded
when calculating the average expectation value of the plaquette.  The
resulting plaquette values, obtained during the two updating
sequences, are shown in Fig.~\ref{hyst}.  No sign of hysteresis is
observed: the algorithm evolves the system to the correct equilibrium
state in a reasonable number of steps, independently of the starting
configuration.

This absence of evidence for metastability, together with the good
performance of our algorithm in all of our production runs, gives us
confidence that our choice of algorithm and of action is appropriate
for the range of parameters that we have considered so far.

\section{Topology}


In phenomenological applications, the combined choice of an action and
an algorithm must allow for an adequate sampling of sectors of
different topological charge. Quite generically, this sampling becomes
more difficult as the continuum limit is approached. Thus, as the
lattice spacing is reduced, the autocorrolation time of topological
charge increases. This is also the case in our simulations. However,
within the range of lattice spacings which we consider, we observe no
dramatic slowing down of tunneling events.

To determine the topological charge of our configurations, we use the
naive gluonic charge definition
\begin{equation}
\label{eq:topch}
q_\mathrm{nai}= \frac{1}{16\pi^2}\sum\limits_{x}\mathrm{Tr}\big[F_{\mu\nu}(x)\tilde{F}^{\mu\nu}(x)\big],
\end{equation}
where $F_{\mu\nu}$ is the gluonic field strength tensor and the sum
extends over all lattice sites. We calculate $F_{\mu\nu}$ at each
lattice site as follows. After applying our smearing prescription (\ref{eq:smearing})
to the links, we average the four plaquettes emanating from this site
and which lie in the $\mu$-$\nu$ plane. The field strength tensor is
then defined as the anti-hermitian part of this average. The charge
defined in Eq.~(\ref{eq:topch}) leads to non-integer values and must be
renormalized for quantative studies of topology. However, such a
renormalization is not necessary here since we are only interested in
verifying the topological ergodicity of our simulations.

\begin{figure}
\begin{center}
\epsfig{file=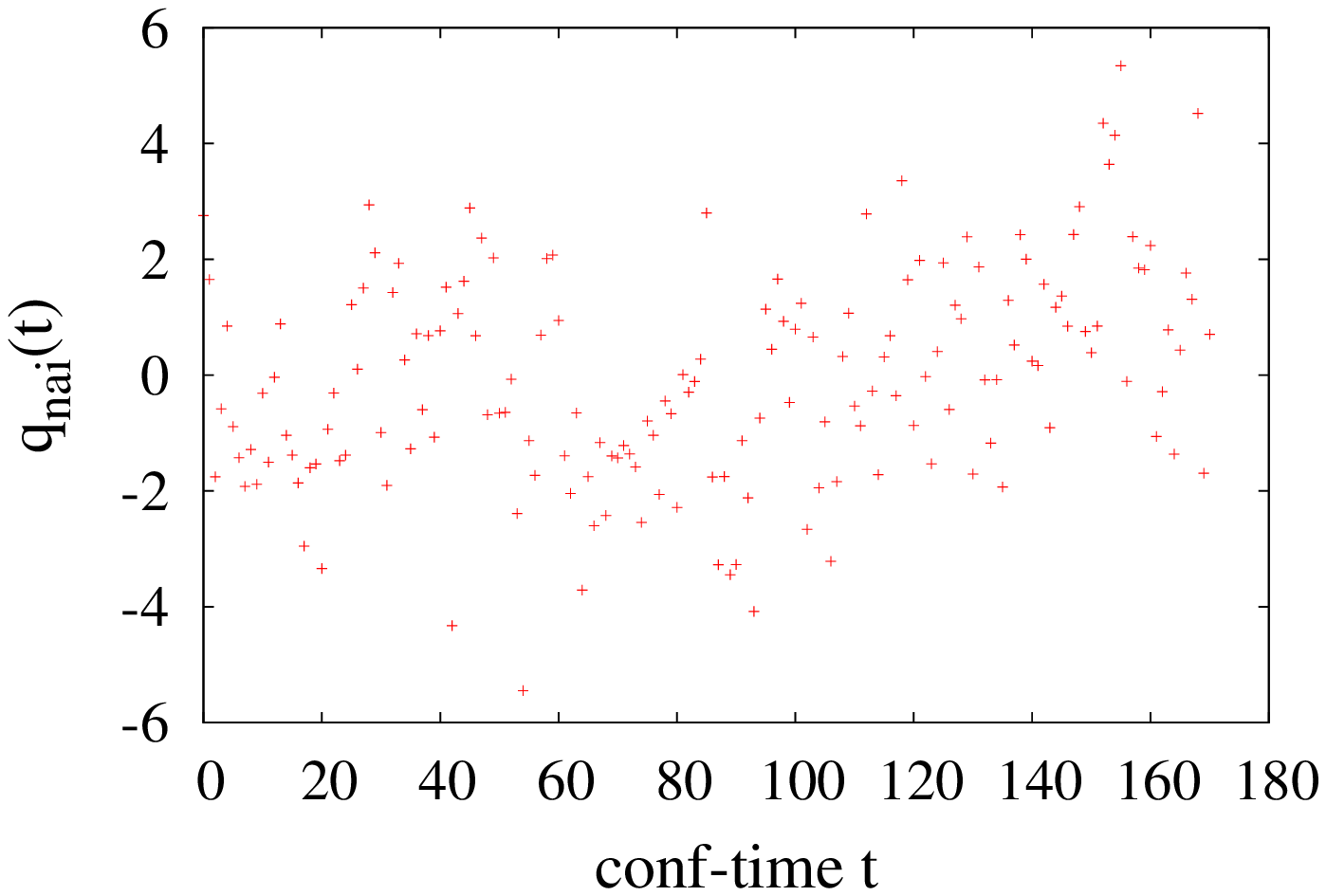,scale=0.55}
\epsfig{file=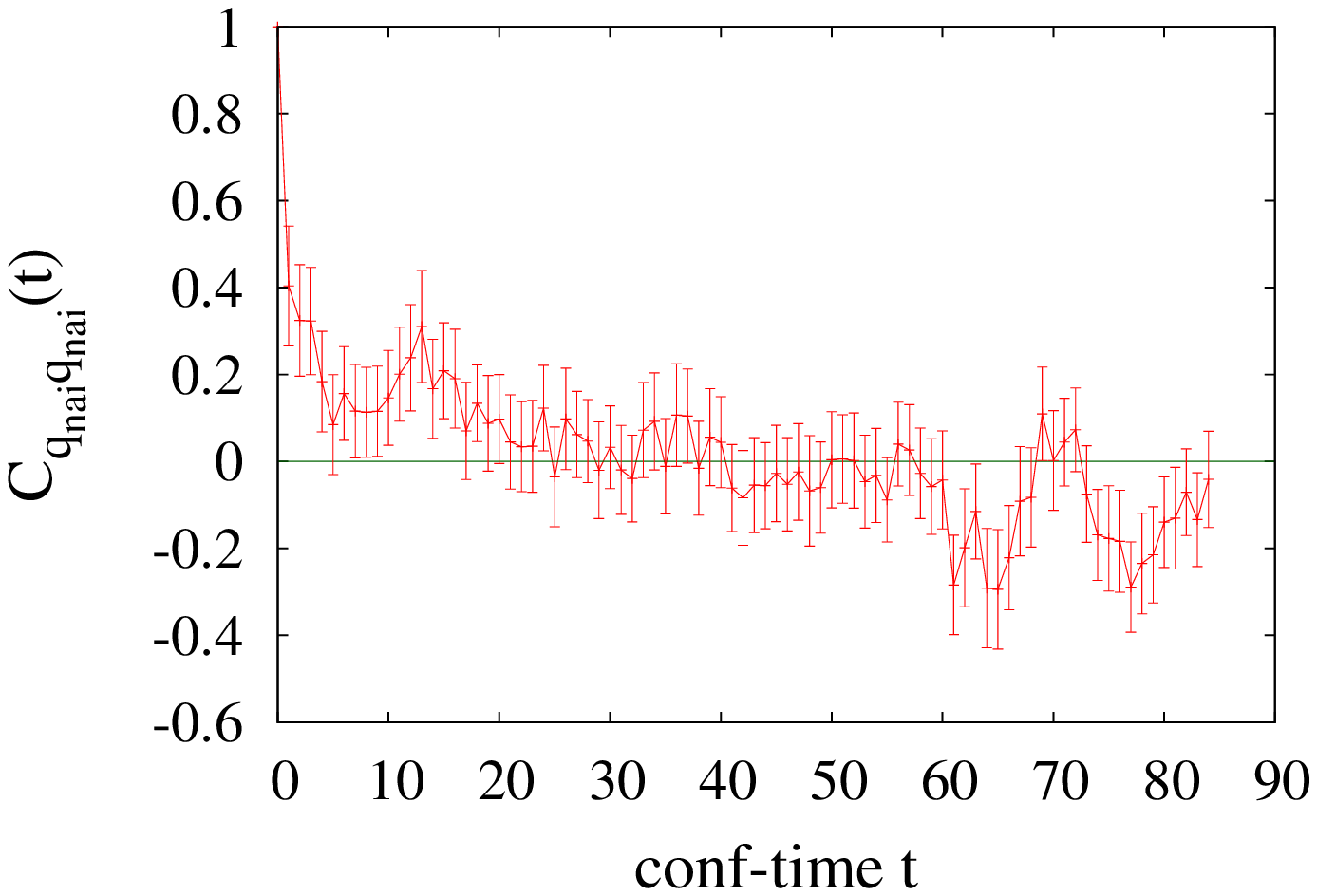,scale=0.55}
\caption{History of the unrenormalized gluonic topological charge (left) and
the corresponding autocorrelation function plot (right), measured on our finest lattice with the
smallest quark mass: $\beta=3.76, aM_\pi=0.2019(20)$. The integrated
autocorrelation time of $q_\mathrm{nai}$ is approximately $2$ configurations on this
ensemble. A separation of one configuration corresponds to 10 HMC/RHMC
trajectories. \label{fig:topo}}
\end{center}
\end{figure}

The simulation-time evolution and autocorrelation of this
unrenormalized topological charge are shown in Fig.\,\ref{fig:topo}
for our finest lattice and its smallest quark mass,
$aM_\pi=0.2019(20)$. The integrated autocorrelation time is around $2$
configurations. The autocorrelation decays very rapidly and is
compatible with zero within the error bars after around 5
configurations.  We can easily conclude from these two plots that
there is no long-range correlation.

\section{Scaling study}



For our scaling study, we use lattices with approximately constant
physical volume at five different lattice spacings. We opted for an
$N_f=3$ instead of an $N_f=2$ setting in order to test the full RHMC
algorithm that is also being used for phenomenological
applications. We choose a $T=2L$ geometry with lattice sizes varying
from $L/a=8$ to $L/a=24$ and bare gauge couplings between $\beta=2.8$
and $\beta=3$.76. We measure fermionic observables every twenty trajectories for
$L/a=8,10,12$ and every ten for $L/a=16,24$. For the error analysis, we use the ``moving-block-bootstrap'' \cite{Mignani:1995}
technique with a binlength of two times the integrated autocorrelation time of the quantity which is measured. This binlength
is typically around $2$ for the coarsest lattices and around $8$ for the finest lattices.
The number of bootstrap samples is chosen to be $2000$, because the calculated bootstrap errors saturate at $\simeq 1500$ samples.

\begin{figure}
\center
\epsfig{file=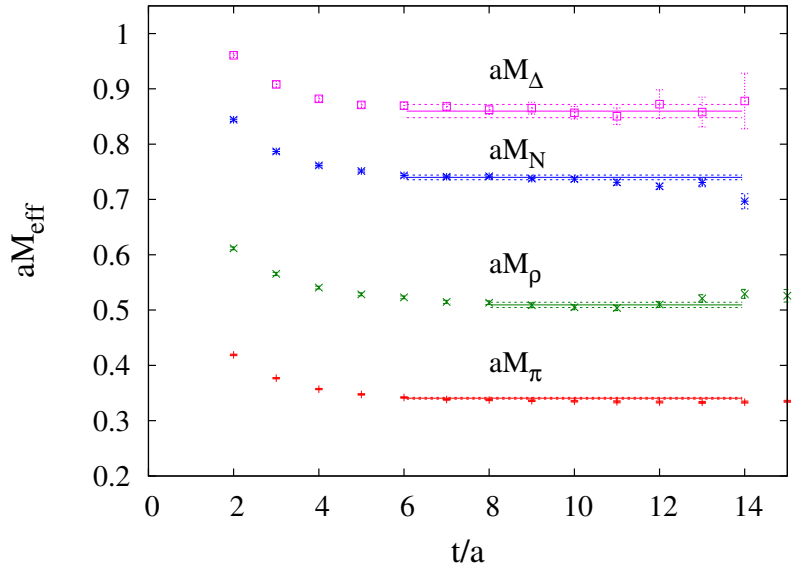,scale=0.9}
\vspace*{-2mm}
\caption{Effective masses of the pion, rho, nucleon and delta on our ensemble
with $L/a = 16$, $\beta = 3.59$, $am_{\mathrm{PCAC}} = 0.04608(12)$. The points
are obtained by solving, for $aM_{\mathrm{eff}}$, the equation $C(t - 1)/C(t + 1) =
f(aM_{\mathrm{eff}}(t - 1 - T/2))/f(aM_{\mathrm{eff}}(t + 1 - T/2))$ at each $t$, where $f(x)=
\cosh(x)$ (for $\pi$ and $\rho$) or $f(x)=\sinh(x)$ (for $N$ and
$\Delta$).  The horizontal lines are the masses with error bars
obtained from correlated cosh or sinh fits to the corresponding
two-point functions in the time intervals indicated by the length of
the lines.\label{fig:effmass}}
\end{figure}

\begin{figure}
\center
\epsfig{file=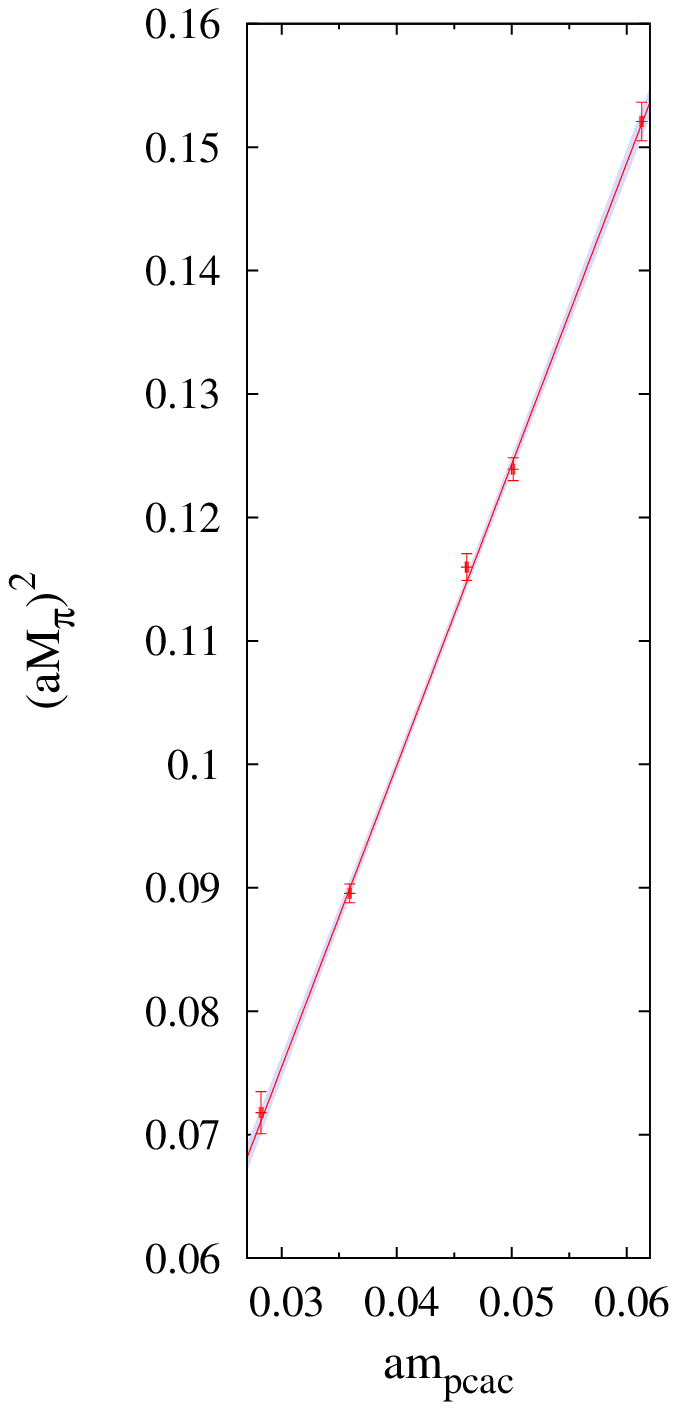,scale=0.8}
\epsfig{file=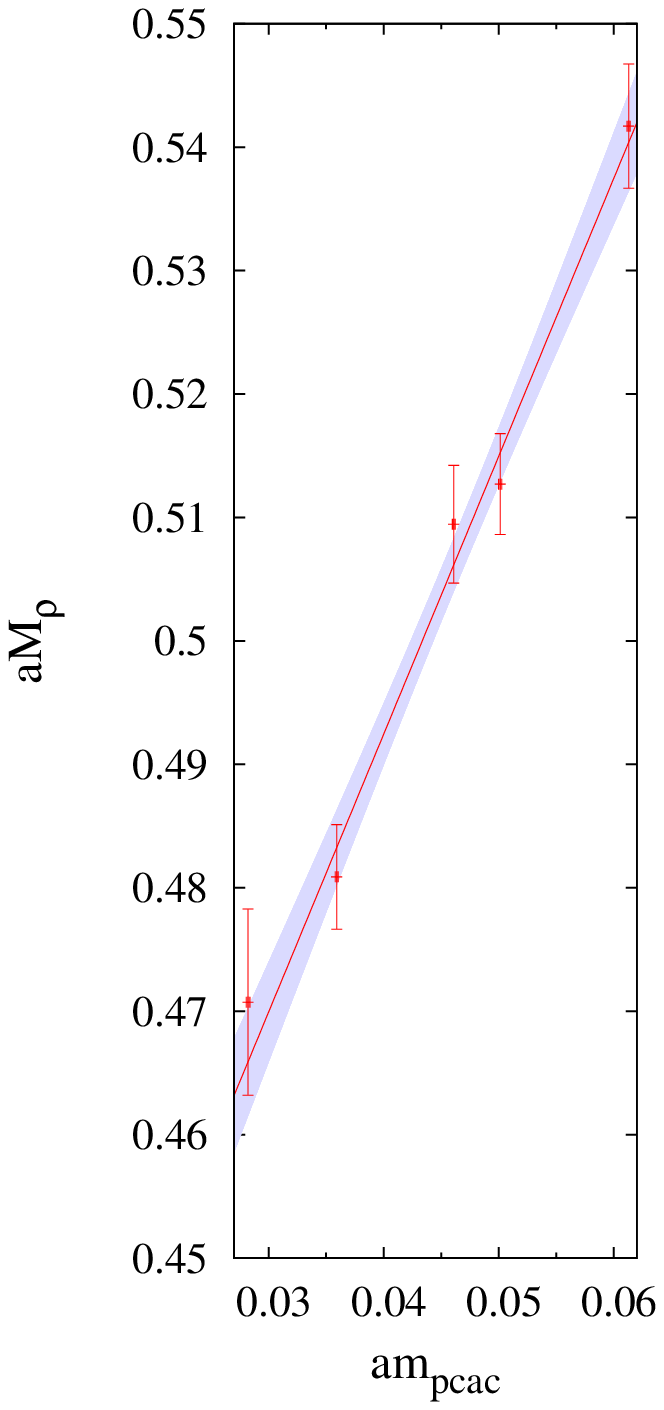,scale=0.8}
\epsfig{file=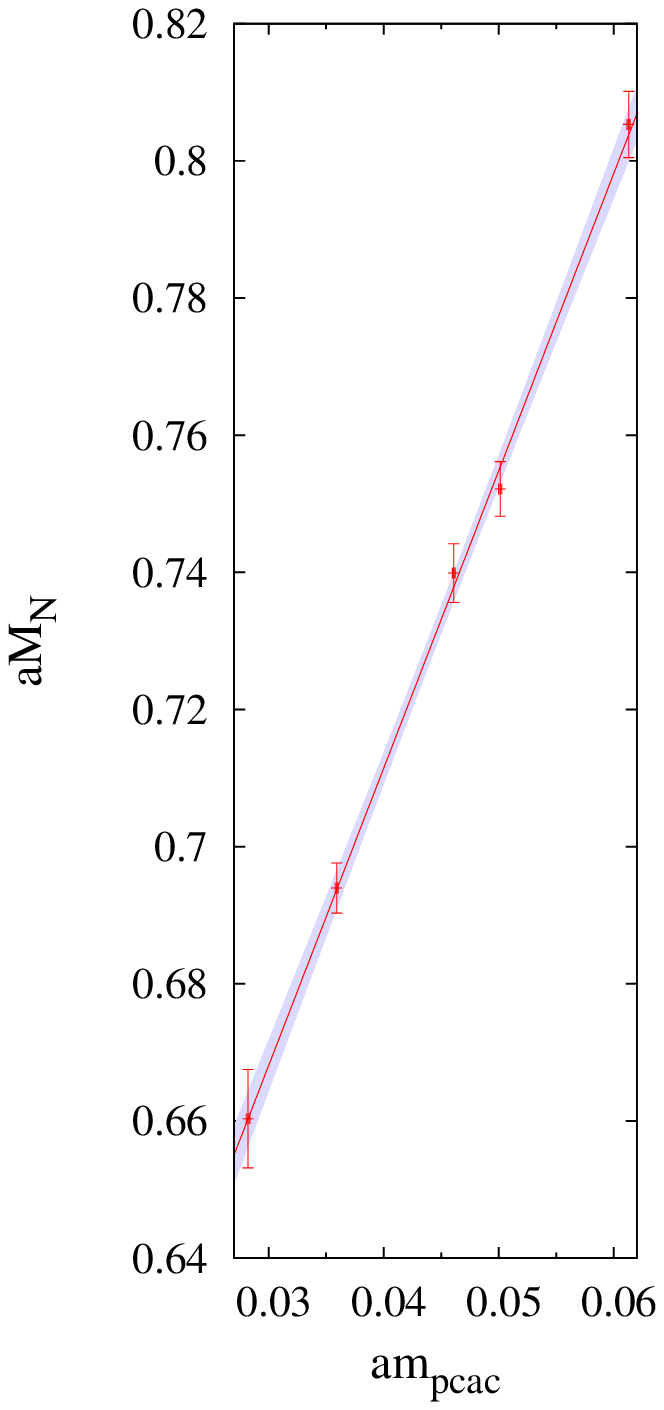,scale=0.8}
\epsfig{file=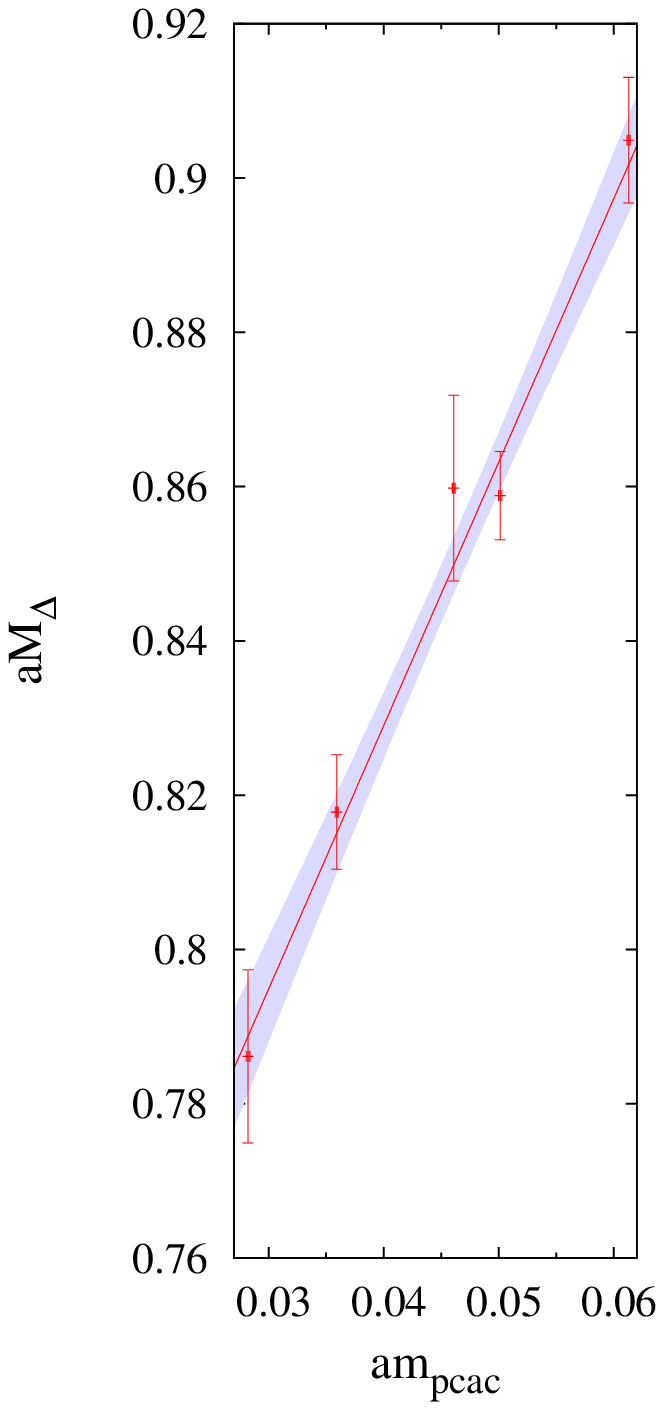,scale=0.8}
\vspace*{-2mm}
\caption{Linear fits of the spectrum in terms of the PCAC quark
  mass. Data shown are from $\beta=3.59$, $L/a=16$ simulations. The line
  indicates the central value of the interpolation and the shaded region is the corresponding
  $1\sigma$ error band (based on the assumption that the linear ansatz
  is correct).
\label{fig:inter}}
\end{figure}

At each lattice spacing we simulate a number of masses (from seven at
$L/a=8$ to three at $L/a=24$) such that $M_\pi/M_\rho$ is between $0.60$
and $0.68$. As already mentioned in the Introduction, it is preferable
to use these rather large masses for a scaling study in order to
enhance possible discretization effects of order $Ma$. After fixing to
Coulomb gauge, we measure propagators with multiple Gaussian sources
on different time slices. The source size is set to $L/4$ and is thus
roughly constant in physical units. Using a Gaussian sink of the same size, the effective
masses usually reach a plateau very quickly and we can determine a useful fitting window from it.
Note that -- because the $x$-space (ultra-)locality of our action is the same as for the unsmeared 
clover action -- such a ``normal'' behavior is exactly what one would expect.
To illustrate this point, a typical effective mass plot is shown in Fig.\,\ref{fig:effmass}.
Then, the masses are extracted from a correlated single channel $\cosh$ or $\sinh$ fit to the correlators.
In order to estimate the systematic error due to excited states, we reduced the initial fit time by up to $2$
timeslices and repeated the analysis with the new fit ranges. This difference then propagates into the systematic error in the continuum limit.

For each coupling $\beta$ we then interpolate $a^2M_\pi^2$, $aM_\rho$,
$aM_{\rm N}$ and $aM_\Delta$ linearly to a common current quark mass
as determined by $M_\pi/M_\rho$.  For illustration the interpolation
at $\beta=3.59$ is shown in Fig.~\ref{fig:inter}.  The error on the
current quark mass is of order $10^{-4}$ and therefore barely visible on this
scale.  Note that all data points are fully unquenched.


\begin{table}
\begin{center}
{\small
\def\arraystretch{1.1}
\begin{tabular}{|c|cc|ll|llll|}
\hline
$M_\pi/M_\rho$ & $\ L/a\ $ & $\ \beta\ $ & $am_{\rm PCAC}$ & $LM_\pi$ & $aM_\pi$ & $aM_\rho$ & $aM_{\rm N}$ & $aM_\Delta$\\
\hline
\multirow{5}{*}{$0.60$} 
& $8$  & $2.80$  & $0.0676(11)$  & $4.55$ & $0.5688(26)$ & $0.9480(44)$ & $1.3605(73)$ & $1.5944(75)$\\
& $10$ & $3.23$  & $0.0468(28)$  & $4.44$ & $0.4437(57)$ & $0.7395(95)$ & $1.064(12)$ & $1.248(10)$\\
& $12$ & $3.40$  & $0.0437(15)$  & $4.60$ & $0.3830(34)$ & $0.6384(57)$ & $0.9236(74)$ & $1.0823(87)$\\
& $16$ & $3.59$  & $0.0328(6)$   & $4.56$ & $0.2852(26)$ & $0.4754(43)$ & $0.6785(44)$ & $0.8031(38)$\\
& $24$ & $3.76$  & $0.0217(7)$   & $4.85$ & $0.2019(20)$ & $0.3365(33)$ & $0.4825(34)$ & $0.5708(20)$\\\hline

\multirow{5}{*}{$0.64$} 
& $8$  & $2.80$  & $0.0839(8)$  & $5.03$  & $0.6292(21)$ & $0.9832(33)$ & $1.4341(43)$ & $1.6581(59)$\\
& $10$ & $3.23$  & $0.0607(23)$ & $4.95$  & $0.4950(47)$ & $0.7735(73)$ & $1.127(10)$ & $1.3074(82)$\\
& $12$ & $3.40$  & $0.0545(13)$ & $5.12$  & $0.4268(23)$ & $0.6669(35)$ & $0.9711(62)$ & $1.1282(71)$\\
& $16$ & $3.59$  & $0.0405(6)$  & $5.03$  & $0.3146(23)$ & $0.4916(36)$ & $0.7099(35)$ & $0.8278(28)$\\
& $24$ & $3.76$  & $0.0270(6)$  & $5.41$  & $0.2256(18)$ & $0.3524(28)$ & $0.5081(29)$ & $0.5933(29)$\\\hline

\multirow{5}{*}{$0.68$} 
& $8$  & $2.80$  & $0.1050(11)$ & $5.60$  & $0.6993(22)$ & $1.0284(32)$ & $1.5286(52)$ & $1.7401(65)$\\
& $10$ & $3.23$  & $0.0796(21)$ & $5.57$  & $0.5574(52)$ & $0.8198(76)$ & $1.212(11)$ & $1.389(10)$\\
& $12$ & $3.40$  & $0.0693(12)$ & $5.76$  & $0.4798(30)$ & $0.7055(44)$ & $1.0354(47)$ & $1.1903(52)$\\
& $16$ & $3.59$  & $0.0506(7)$  & $5.57$  & $0.3483(22)$ & $0.5122(32)$ & $0.7495(30)$ & $0.8590(41)$\\
& $24$ & $3.76$  & $0.0343(9)$  & $6.11$  & $0.2546(25)$ & $0.3744(37)$ & $0.5434(38)$ & $0.6242(39)$\\\hline
\end{tabular}}
\def\arraystretch{1.0}
\end{center}
\vspace*{-4mm}
\caption{Results of the interpolation of $aM_\pi$, $aM_\rho$, $aM_N$
  and $aM_\Delta$, obtained from simulations performed at different
  bare quark masses and gauge couplings, to the reference points
  $M_\pi/M_\rho=0.60,0.64,0.68$. \label{tab:inter}}
\end{table}

We perform our scaling test on the baryon spectrum for three different
values of $M_\pi/M_\rho$, all of which can be reached by interpolating
our simulation data.  In Tab.\,\ref{tab:inter} we summarize the values
of $am_{\rm PCAC}$, $aM_\pi$, $aM_\rho$, $aM_{\rm N}$ and $aM_\Delta$
after interpolation to $M_\pi/M_\rho=0.60,0.64,0.68$.  Also listed is
$LM_\pi$, which is roughly constant for fixed $M_\pi/M_\rho$.
Moreover, even for the lightest data set we are deep in the $M_\pi
L\!>\!4$ regime.
In case this criterion alone would not garantee the smallness of finite
volume effects, the fact that our boxes have a fixed physical size
ensures that such effects would be the same for all data at a given
$M_\pi/M_\rho$ ratio, and the scaling test would still be meaningful.

The masses are known to better than $2\%$ and, due to correlations,
this is also true for mass ratios. For the three lines of constant
physics, $M_{\rm N}$ and $M_\Delta$ in units of $M_\pi$ are plotted in
Fig.\,\ref{fig:bary} as functions of the squared lattice spacing (see
below), measured in units of the vector meson mass. We normalize the baryon
masses by $M_\pi$ to clearly separate the lines of constant physics in the plot. 
The fits incorporate the error bars along both the vertical and horizontal axes.

For both the spin-$1/2$ and spin-$3/2$ baryons, the continuum limit is
approached smoothly with scaling violations of at most 1.2\% at
$\beta=2.8$. The extrapolations shown exclude this data
point but consistent results are obtained by using all available data.

While we expect that our choice of the clover coefficient is close to
a non-perturbatively determined value, we cannot exclude effects that
are linear in the lattice spacing in principle. The cutoff effects
that we consider here are so small that we can not make a definitive
statement, despite the fact that we have very precise data and cover
more than a factor of seven in $a^2$.  Assuming the lattice artifacts
to be linear in $a$ results in an only marginally worse fit.

An alternative way of proceeding is doing a combined chiral and 
continuum extrapolation with all datapoints at once. Applying this procedure one obtains 
basically consistent continuum limits and we assume
the absolute differences as our systematic errors.

\begin{figure}
\center
\epsfig{file=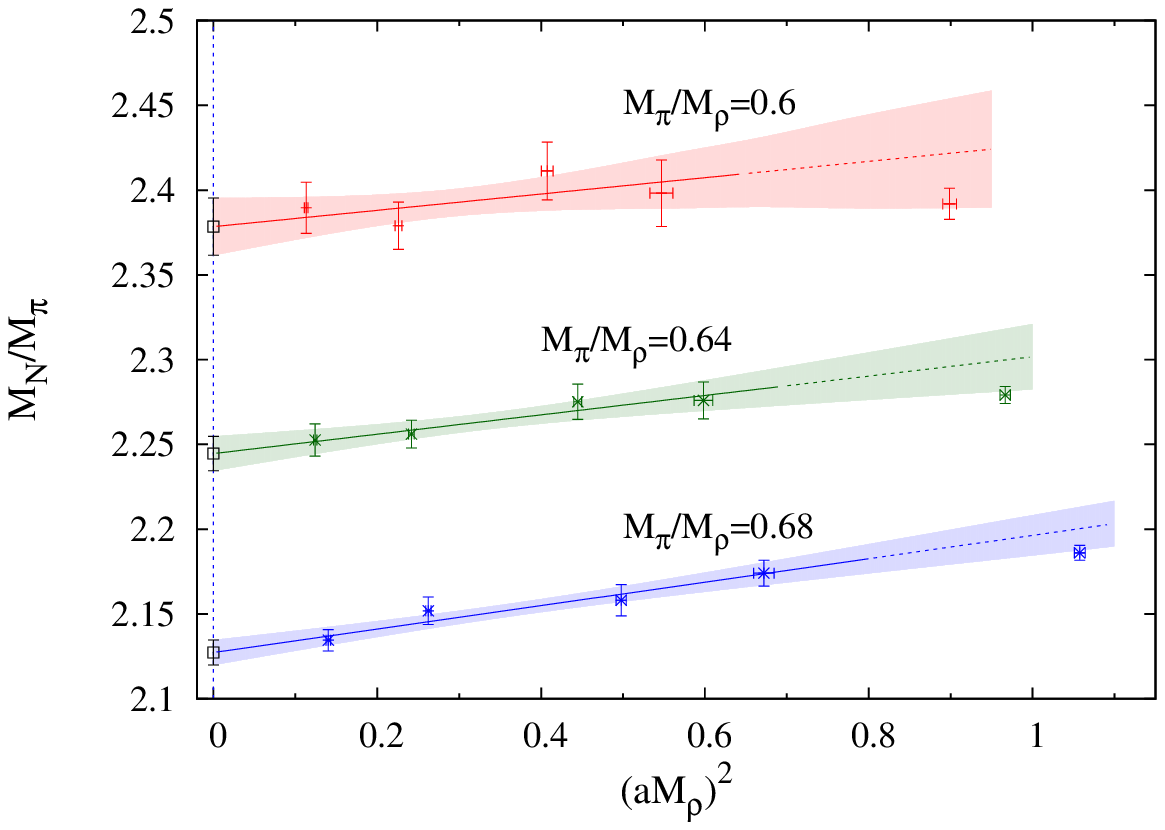,scale=1.2}
\hspace{-2mm}
\epsfig{file=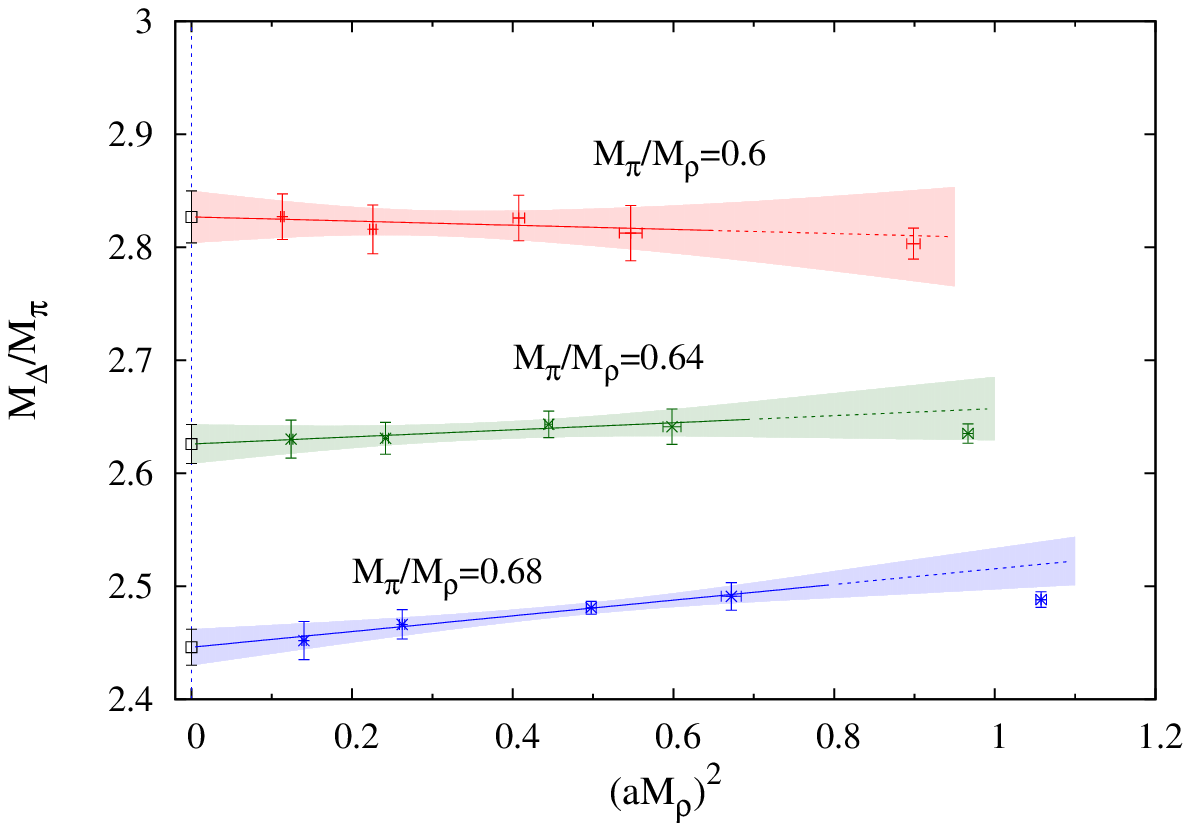,scale=1.2}
\vspace*{-4mm}
\caption{$M_{\rm N}$ and $M_\Delta$, the mass of the spin-$1/2$ and spin-$3/2$
baryon, in terms of $M_\pi$, versus the lattice spacing squared (in terms of
$M_\rho^{-1}$). Each one of the three continuum extrapolations is based on the
data at $\beta=3.76-3.23$, but the curve is extended to $\beta=2.8$ to allow for
comparison. The continuum limits are $M_N/M_\pi=2.378(17)(43),\,2.245(10)(51),\,2.127(7)(34)$ 
and $M_\Delta/M_\pi=2.827(23)(40),\,2.626(17)(49),\,2.446(16)(30)$ respectively. For all datapoints only statistical errors
are shown.\label{fig:bary}}
\end{figure}

\begin{figure}
\center
\epsfig{file=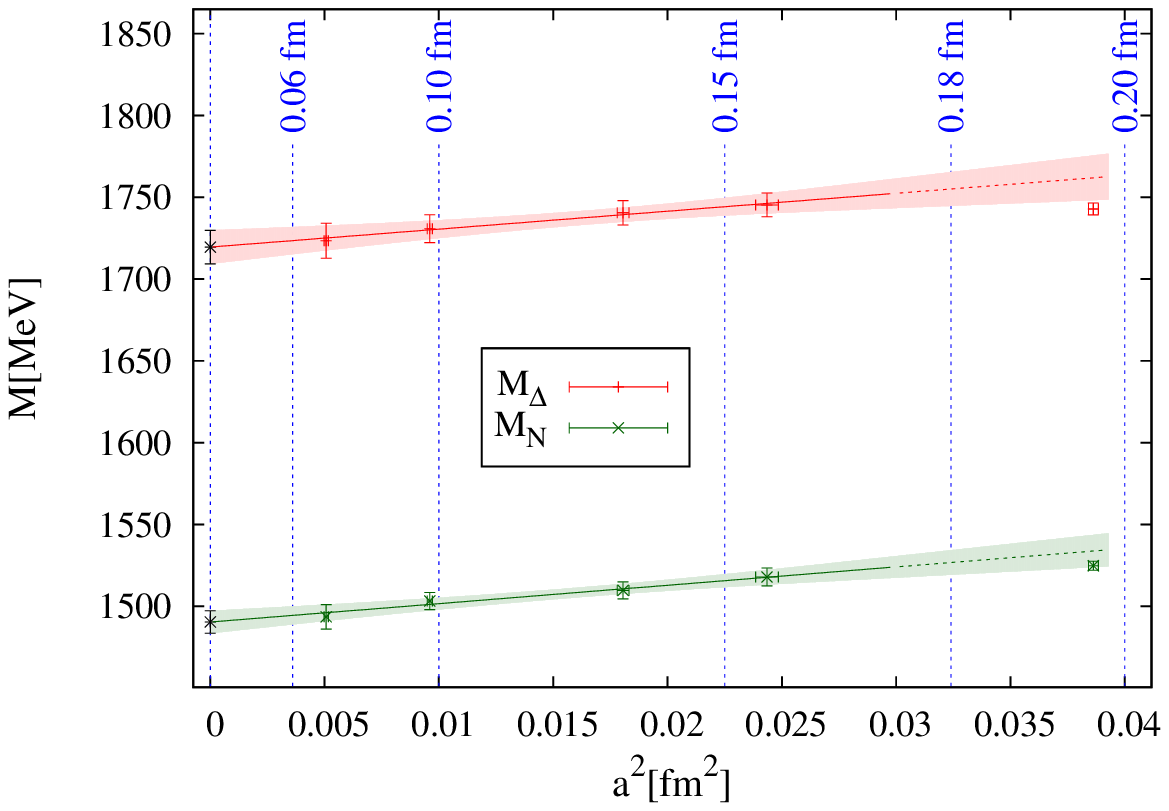,scale=1.2}
\vspace*{-2mm}
\caption{
\label{fig:physical}
Scaling of the $\Delta$ and nucleon mass at $M_\pi/ M_\rho=0.67$ in
physical units using the scale setting procedure described around
(\ref{eq:scale}). In the continuum we obtain
$M_\mathrm{N}=1490(7)(27)$~MeV and $M_\Delta=1720(10)(35)$~MeV. As in Fig.\,\ref{fig:bary}, only statistical errors are shown.}
\end{figure}

For illustrative purposes, we set the scale by linearly interpolating
 $M_\rho$ and $M_\pi^2$ to the point where
\begin{equation}
{M_\pi/
 M_\rho}={\sqrt{2(M_K^{\mathrm{phys}})^2-(M_\pi^{\mathrm{phys}})^2}/
 M_\phi^{\mathrm{phys}}}\sim 0.67
\label{eq:scale}
\end{equation}
and identify $M_\rho$ with the mass of the physical $\phi$. In this
convention we cover lattice spacings from about $0.19$~fm down to
$0.07$~fm (see Fig.~\ref{fig:physical}). In this range we find only
small scaling violations in the spectrum and those disappear smoothly
toward the continuum. The behavior is consistent with that of an ${\rm
O}(a)$-improved theory.

The scaling of other observables, especially matrix elements, will be
investigated in the future.


\section{Summary}


We have described an efficient algorithm to perform full lattice QCD
calculations with stout-link, improved clover fermions and
demonstrated its potential with a scaling study of light baryon masses
in $N_f=3$ QCD. We have tested the algorithm and found it to be
stable and reliable down to relatively coarse lattices with $a\simeq 0.16$~fm. 
We have also monitored the stability of the MD
integration and the lowest eigenvalue of the (even-odd preconditioned)
fermion matrix and demonstrated that the latter is sufficiently far
away from zero on all of our ensembles. Furthermore, we have shown that
there is no sign of exceptional configurations even with substantially
lighter pion masses in an $N_f=2+1$ setting.

Upon performing a ``thermal cycle'' at $\beta=3.3$, with $M_\pi$
ranging between $\sim240$~MeV and $\sim440$~MeV, we do not see any sign of a
hysteresis. In other words, there is no indication of a nearby first
order phase transition, even on fairly coarse lattices and for rather
light quark masses.

In a dedicated scaling test of light baryon masses, which included
five lattice spacings with a total variation by almost a factor of
three, we have demonstrated that scaling violations associated with
the use of our stout-link clover action in full QCD are small for
these quantities. Indeed, we have shown that discretization errors on
light baryon masses do not exceed $2\%$ for lattice spacings up to
$0.19$~fm. Moreover, all our data for $a\leq 0.16$~fm seem to be in the scaling window. 
This is in line with the findings of~\cite{Kamleh:2007bd}
where a different approach to link smearing is taken.

In conclusion, we find that the combination of a tree-level Symanzik
improved gauge action and a six-step stout-smeared clover fermion
action with $c_\mathrm{SW}=1$ is well suited for precision
calculations of physical observables. We expect that the same will be
true of other actions with comparable improvements. We look forward to
presenting results for phenomenological quantities with this action in
forthcoming papers.

\acknowledgments Computations were performed on the BlueGene/P at FZ
J{\"u}lich and on clusters at the University of Wuppertal and at CPT
Marseille. This work is supported in part by EU grant I3HP, OTKA
grants T34980,T37615,M37071,T032501,AT049652, DFG grant FO 502/1-2, EU
RTN contract MRTN-CT-2006-035482 (FLAVIAnet) and by the CNRS's GDR
grant n$^o$ 2921 (``Physique subatomique et calculs sur r\'eseau'').

\bibliography{refs}

\begin{thebibliography}{35}
\expandafter\ifx\csname natexlab\endcsname\relax\def\natexlab#1{#1}\fi
\expandafter\ifx\csname bibnamefont\endcsname\relax
  \def\bibnamefont#1{#1}\fi
\expandafter\ifx\csname bibfnamefont\endcsname\relax
  \def\bibfnamefont#1{#1}\fi
\expandafter\ifx\csname citenamefont\endcsname\relax
  \def\citenamefont#1{#1}\fi
\expandafter\ifx\csname url\endcsname\relax
  \def\url#1{\texttt{#1}}\fi
\expandafter\ifx\csname urlprefix\endcsname\relax\def\urlprefix{URL }\fi
\providecommand{\bibinfo}[2]{#2}
\providecommand{\eprint}[2][]{\url{#2}}

\bibitem[{\citenamefont{Albanese et~al.}(1987)}]{Albanese:1987ds}
\bibinfo{author}{\bibfnamefont{M.}~\bibnamefont{Albanese}} \bibnamefont{et~al.}
  (\bibinfo{collaboration}{APE}), \bibinfo{journal}{Phys. Lett.}
  \textbf{\bibinfo{volume}{B192}}, \bibinfo{pages}{163} (\bibinfo{year}{1987}).

\bibitem[{\citenamefont{Sheikholeslami and
  Wohlert}(1985)}]{Sheikholeslami:1985ij}
\bibinfo{author}{\bibfnamefont{B.}~\bibnamefont{Sheikholeslami}}
  \bibnamefont{and} \bibinfo{author}{\bibfnamefont{R.}~\bibnamefont{Wohlert}},
  \bibinfo{journal}{Nucl. Phys.} \textbf{\bibinfo{volume}{B259}},
  \bibinfo{pages}{572} (\bibinfo{year}{1985}).

\bibitem[{\citenamefont{DeGrand et~al.}(1998)\citenamefont{DeGrand, Hasenfratz,
  and Kovacs}}]{DeGrand:1998jq}
\bibinfo{author}{\bibfnamefont{T.~A.} \bibnamefont{DeGrand}},
  \bibinfo{author}{\bibfnamefont{A.}~\bibnamefont{Hasenfratz}},
  \bibnamefont{and} \bibinfo{author}{\bibfnamefont{T.~G.} \bibnamefont{Kovacs}}
  (\bibinfo{collaboration}{MILC}) (\bibinfo{year}{1998}),
  \eprint{hep-lat/9807002}.

\bibitem[{\citenamefont{Bernard and DeGrand}(2000)}]{Bernard:1999kc}
\bibinfo{author}{\bibfnamefont{C.~W.} \bibnamefont{Bernard}} \bibnamefont{and}
  \bibinfo{author}{\bibfnamefont{T.~A.} \bibnamefont{DeGrand}},
  \bibinfo{journal}{Nucl. Phys. Proc. Suppl.} \textbf{\bibinfo{volume}{83}},
  \bibinfo{pages}{845} (\bibinfo{year}{2000}), \eprint{hep-lat/9909083}.

\bibitem[{\citenamefont{Stephenson et~al.}(2001)\citenamefont{Stephenson,
  Detar, DeGrand, and Hasenfratz}}]{Stephenson:1999ns}
\bibinfo{author}{\bibfnamefont{M.}~\bibnamefont{Stephenson}},
  \bibinfo{author}{\bibfnamefont{C.~E.} \bibnamefont{Detar}},
  \bibinfo{author}{\bibfnamefont{T.~A.} \bibnamefont{DeGrand}},
  \bibnamefont{and}
  \bibinfo{author}{\bibfnamefont{A.}~\bibnamefont{Hasenfratz}},
  \bibinfo{journal}{Phys. Rev.} \textbf{\bibinfo{volume}{D63}},
  \bibinfo{pages}{034501} (\bibinfo{year}{2001}), \eprint{hep-lat/9910023}.

\bibitem[{\citenamefont{Bernard et~al.}(2001)}]{Bernard:2000ht}
\bibinfo{author}{\bibfnamefont{C.~W.} \bibnamefont{Bernard}}
  \bibnamefont{et~al.}, \bibinfo{journal}{Nucl. Phys. Proc. Suppl.}
  \textbf{\bibinfo{volume}{94}}, \bibinfo{pages}{346} (\bibinfo{year}{2001}),
  \eprint{hep-lat/0011029}.

\bibitem[{\citenamefont{Zanotti et~al.}(2002)}]{Zanotti:2001yb}
\bibinfo{author}{\bibfnamefont{J.~M.} \bibnamefont{Zanotti}}
  \bibnamefont{et~al.} (\bibinfo{collaboration}{CSSM Lattice}),
  \bibinfo{journal}{Phys. Rev.} \textbf{\bibinfo{volume}{D65}},
  \bibinfo{pages}{074507} (\bibinfo{year}{2002}), \eprint{hep-lat/0110216}.

\bibitem[{\citenamefont{Capitani et~al.}(2006)\citenamefont{Capitani, D{\"u}rr,
  and Hoelbling}}]{Capitani:2006ni}
\bibinfo{author}{\bibfnamefont{S.}~\bibnamefont{Capitani}},
  \bibinfo{author}{\bibfnamefont{S.}~\bibnamefont{D{\"u}rr}}, \bibnamefont{and}
  \bibinfo{author}{\bibfnamefont{C.}~\bibnamefont{Hoelbling}},
  \bibinfo{journal}{JHEP} \textbf{\bibinfo{volume}{11}}, \bibinfo{pages}{028}
  (\bibinfo{year}{2006}), \eprint{hep-lat/0607006}.

\bibitem[{\citenamefont{Morningstar and Peardon}(2004)}]{Morningstar:2003gk}
\bibinfo{author}{\bibfnamefont{C.}~\bibnamefont{Morningstar}} \bibnamefont{and}
  \bibinfo{author}{\bibfnamefont{M.~J.} \bibnamefont{Peardon}},
  \bibinfo{journal}{Phys. Rev.} \textbf{\bibinfo{volume}{D69}},
  \bibinfo{pages}{054501} (\bibinfo{year}{2004}), \eprint{hep-lat/0311018}.

\bibitem[{\citenamefont{Duane et~al.}(1987)\citenamefont{Duane, Kennedy,
  Pendleton, and Roweth}}]{Duane:1987de}
\bibinfo{author}{\bibfnamefont{S.}~\bibnamefont{Duane}},
  \bibinfo{author}{\bibfnamefont{A.~D.} \bibnamefont{Kennedy}},
  \bibinfo{author}{\bibfnamefont{B.~J.} \bibnamefont{Pendleton}},
  \bibnamefont{and} \bibinfo{author}{\bibfnamefont{D.}~\bibnamefont{Roweth}},
  \bibinfo{journal}{Phys. Lett.} \textbf{\bibinfo{volume}{B195}},
  \bibinfo{pages}{216} (\bibinfo{year}{1987}).

\bibitem[{\citenamefont{Kamleh et~al.}(2004)\citenamefont{Kamleh, Leinweber,
  and Williams}}]{Kamleh:2004xk}
\bibinfo{author}{\bibfnamefont{W.}~\bibnamefont{Kamleh}},
  \bibinfo{author}{\bibfnamefont{D.~B.} \bibnamefont{Leinweber}},
  \bibnamefont{and} \bibinfo{author}{\bibfnamefont{A.~G.}
  \bibnamefont{Williams}}, \bibinfo{journal}{Phys. Rev.}
  \textbf{\bibinfo{volume}{D70}}, \bibinfo{pages}{014502}
  (\bibinfo{year}{2004}), \eprint{hep-lat/0403019}.

\bibitem[{\citenamefont{Hasenfratz et~al.}(2007)\citenamefont{Hasenfratz,
  Hoffmann, and Schaefer}}]{Hasenfratz:2007rf}
\bibinfo{author}{\bibfnamefont{A.}~\bibnamefont{Hasenfratz}},
  \bibinfo{author}{\bibfnamefont{R.}~\bibnamefont{Hoffmann}}, \bibnamefont{and}
  \bibinfo{author}{\bibfnamefont{S.}~\bibnamefont{Schaefer}},
  \bibinfo{journal}{JHEP} \textbf{\bibinfo{volume}{05}}, \bibinfo{pages}{029}
  (\bibinfo{year}{2007}), \eprint{hep-lat/0702028}.

\bibitem[{\citenamefont{D{\"u}rr}(2007)}]{Durr:2007cy}
\bibinfo{author}{\bibfnamefont{S.}~\bibnamefont{D{\"u}rr}}
  (\bibinfo{year}{2007}), \eprint{arXiv:0709.4110 [hep-lat]}.

\bibitem[{\citenamefont{Schaefer et~al.}(2007)\citenamefont{Schaefer,
  Hasenfratz, and Hoffmann}}]{Schaefer:2007dc}
\bibinfo{author}{\bibfnamefont{S.}~\bibnamefont{Schaefer}},
  \bibinfo{author}{\bibfnamefont{A.}~\bibnamefont{Hasenfratz}},
  \bibnamefont{and} \bibinfo{author}{\bibfnamefont{R.}~\bibnamefont{Hoffmann}},
  \bibinfo{journal}{PoS} \textbf{\bibinfo{volume}{LAT2007}},
  \bibinfo{pages}{132} (\bibinfo{year}{2007}), \eprint{arXiv:0709.4130
  [hep-lat]}.

\bibitem[{\citenamefont{Hoffmann et~al.}(2007)\citenamefont{Hoffmann,
  Hasenfratz, and Schaefer}}]{Hoffmann:2007nm}
\bibinfo{author}{\bibfnamefont{R.}~\bibnamefont{Hoffmann}},
  \bibinfo{author}{\bibfnamefont{A.}~\bibnamefont{Hasenfratz}},
  \bibnamefont{and} \bibinfo{author}{\bibfnamefont{S.}~\bibnamefont{Schaefer}},
  \bibinfo{journal}{PoS} \textbf{\bibinfo{volume}{LAT2007}},
  \bibinfo{pages}{104} (\bibinfo{year}{2007}), \eprint{arXiv:0710.0471
  [hep-lat]}.

\bibitem[{\citenamefont{Moran and Leinweber}(2008)}]{Moran:2008ra}
\bibinfo{author}{\bibfnamefont{P.~J.} \bibnamefont{Moran}} \bibnamefont{and}
  \bibinfo{author}{\bibfnamefont{D.~B.} \bibnamefont{Leinweber}}
  (\bibinfo{year}{2008}), \eprint{arXiv:0801.1165 [hep-lat]}.

\bibitem[{\citenamefont{Della~Morte et~al.}(2005)\citenamefont{Della~Morte,
  Hoffmann, Knechtli, and Wolff}}]{DellaMorte:2004hs}
\bibinfo{author}{\bibfnamefont{M.}~\bibnamefont{Della~Morte}},
  \bibinfo{author}{\bibfnamefont{R.}~\bibnamefont{Hoffmann}},
  \bibinfo{author}{\bibfnamefont{F.}~\bibnamefont{Knechtli}}, \bibnamefont{and}
  \bibinfo{author}{\bibfnamefont{U.}~\bibnamefont{Wolff}}
  (\bibinfo{collaboration}{ALPHA}), \bibinfo{journal}{Comput. Phys. Commun.}
  \textbf{\bibinfo{volume}{165}}, \bibinfo{pages}{49} (\bibinfo{year}{2005}),
  \eprint{hep-lat/0405017}.

\bibitem[{\citenamefont{Aoki et~al.}(2005)}]{Aoki:2004iq}
\bibinfo{author}{\bibfnamefont{S.}~\bibnamefont{Aoki}} \bibnamefont{et~al.}
  (\bibinfo{collaboration}{JLQCD}), \bibinfo{journal}{Phys. Rev.}
  \textbf{\bibinfo{volume}{D72}}, \bibinfo{pages}{054510}
  (\bibinfo{year}{2005}), \eprint{hep-lat/0409016}.

\bibitem[{\citenamefont{L{\"u}scher and Weisz}(1985)}]{Luscher:1985zq}
\bibinfo{author}{\bibfnamefont{M.}~\bibnamefont{L{\"u}scher}} \bibnamefont{and}
  \bibinfo{author}{\bibfnamefont{P.}~\bibnamefont{Weisz}},
  \bibinfo{journal}{Phys. Lett.} \textbf{\bibinfo{volume}{B158}},
  \bibinfo{pages}{250} (\bibinfo{year}{1985}).

\bibitem[{\citenamefont{Wohlert}(1987)}]{Wohlert:1987rf}
\bibinfo{author}{\bibfnamefont{R.}~\bibnamefont{Wohlert}}
  (\bibinfo{year}{1987}), \bibinfo{note}{{DESY 87/069}}.

\bibitem[{\citenamefont{Horsley et~al.}(2007)\citenamefont{Horsley, Perlt,
  Schiller, Rakow, and Schierholz}}]{Horsley:2007fw}
\bibinfo{author}{\bibfnamefont{R.}~\bibnamefont{Horsley}},
  \bibinfo{author}{\bibfnamefont{H.}~\bibnamefont{Perlt}},
  \bibinfo{author}{\bibfnamefont{A.}~\bibnamefont{Schiller}},
  \bibinfo{author}{\bibfnamefont{P.~E.~L.} \bibnamefont{Rakow}},
  \bibnamefont{and}
  \bibinfo{author}{\bibfnamefont{G.}~\bibnamefont{Schierholz}}
  (\bibinfo{year}{2007}), \eprint{arXiv:0710.0990 [hep-lat]}.

\bibitem[{\citenamefont{L{\"u}scher et~al.}(1997)\citenamefont{L{\"u}scher,
  Sint, Sommer, Weisz, and Wolff}}]{Luscher:1996ug}
\bibinfo{author}{\bibfnamefont{M.}~\bibnamefont{L{\"u}scher}},
  \bibinfo{author}{\bibfnamefont{S.}~\bibnamefont{Sint}},
  \bibinfo{author}{\bibfnamefont{R.}~\bibnamefont{Sommer}},
  \bibinfo{author}{\bibfnamefont{P.}~\bibnamefont{Weisz}}, \bibnamefont{and}
  \bibinfo{author}{\bibfnamefont{U.}~\bibnamefont{Wolff}},
  \bibinfo{journal}{Nucl. Phys.} \textbf{\bibinfo{volume}{B491}},
  \bibinfo{pages}{323} (\bibinfo{year}{1997}), \eprint{hep-lat/9609035}.

\bibitem[{\citenamefont{Clark et~al.}(2003)\citenamefont{Clark, Joo, and
  Kennedy}}]{Clark:2002vz}
\bibinfo{author}{\bibfnamefont{M.~A.} \bibnamefont{Clark}},
  \bibinfo{author}{\bibfnamefont{B.}~\bibnamefont{Joo}}, \bibnamefont{and}
  \bibinfo{author}{\bibfnamefont{A.~D.} \bibnamefont{Kennedy}},
  \bibinfo{journal}{Nucl. Phys. Proc. Suppl.} \textbf{\bibinfo{volume}{119}},
  \bibinfo{pages}{1015} (\bibinfo{year}{2003}), \eprint{hep-lat/0209035}.

\bibitem[{\citenamefont{Clark and Kennedy}(2007)}]{Clark:2006fx}
\bibinfo{author}{\bibfnamefont{M.~A.} \bibnamefont{Clark}} \bibnamefont{and}
  \bibinfo{author}{\bibfnamefont{A.~D.} \bibnamefont{Kennedy}},
  \bibinfo{journal}{Phys. Rev. Lett.} \textbf{\bibinfo{volume}{98}},
  \bibinfo{pages}{051601} (\bibinfo{year}{2007}), \eprint{hep-lat/0608015}.

\bibitem[{\citenamefont{DeGrand and Rossi}(1990)}]{DeGrand:1990dk}
\bibinfo{author}{\bibfnamefont{T.~A.} \bibnamefont{DeGrand}} \bibnamefont{and}
  \bibinfo{author}{\bibfnamefont{P.}~\bibnamefont{Rossi}},
  \bibinfo{journal}{Comput. Phys. Commun.} \textbf{\bibinfo{volume}{60}},
  \bibinfo{pages}{211} (\bibinfo{year}{1990}).

\bibitem[{\citenamefont{Ali~Khan et~al.}(2003)}]{AliKhan:2003br}
\bibinfo{author}{\bibfnamefont{A.}~\bibnamefont{Ali~Khan}} \bibnamefont{et~al.}
  (\bibinfo{collaboration}{QCDSF}), \bibinfo{journal}{Phys. Lett.}
  \textbf{\bibinfo{volume}{B564}}, \bibinfo{pages}{235} (\bibinfo{year}{2003}),
  \eprint{hep-lat/0303026}.

\bibitem[{\citenamefont{Urbach et~al.}(2006)\citenamefont{Urbach, Jansen,
  Shindler, and Wenger}}]{Urbach:2005ji}
\bibinfo{author}{\bibfnamefont{C.}~\bibnamefont{Urbach}},
  \bibinfo{author}{\bibfnamefont{K.}~\bibnamefont{Jansen}},
  \bibinfo{author}{\bibfnamefont{A.}~\bibnamefont{Shindler}}, \bibnamefont{and}
  \bibinfo{author}{\bibfnamefont{U.}~\bibnamefont{Wenger}},
  \bibinfo{journal}{Comput. Phys. Commun.} \textbf{\bibinfo{volume}{174}},
  \bibinfo{pages}{87} (\bibinfo{year}{2006}), \eprint{hep-lat/0506011}.

\bibitem[{\citenamefont{Sexton and Weingarten}(1992)}]{Sexton:1992nu}
\bibinfo{author}{\bibfnamefont{J.~C.} \bibnamefont{Sexton}} \bibnamefont{and}
  \bibinfo{author}{\bibfnamefont{D.~H.} \bibnamefont{Weingarten}},
  \bibinfo{journal}{Nucl. Phys.} \textbf{\bibinfo{volume}{B380}},
  \bibinfo{pages}{665} (\bibinfo{year}{1992}).

\bibitem[{\citenamefont{Hasenbusch}(2001)}]{Hasenbusch:2001ne}
\bibinfo{author}{\bibfnamefont{M.}~\bibnamefont{Hasenbusch}},
  \bibinfo{journal}{Phys. Lett.} \textbf{\bibinfo{volume}{B519}},
  \bibinfo{pages}{177} (\bibinfo{year}{2001}), \eprint{hep-lat/0107019}.

\bibitem[{\citenamefont{Takaishi and de~Forcrand}(2006)}]{Takaishi:2005tz}
\bibinfo{author}{\bibfnamefont{T.}~\bibnamefont{Takaishi}} \bibnamefont{and}
  \bibinfo{author}{\bibfnamefont{P.}~\bibnamefont{de~Forcrand}},
  \bibinfo{journal}{Phys. Rev.} \textbf{\bibinfo{volume}{E73}},
  \bibinfo{pages}{036706} (\bibinfo{year}{2006}), \eprint{hep-lat/0505020}.

\bibitem[{\citenamefont{Del~Debbio et~al.}(2006)\citenamefont{Del~Debbio,
  Giusti, L{\"u}scher, Petronzio, and Tantalo}}]{DelDebbio:2005qa}
\bibinfo{author}{\bibfnamefont{L.}~\bibnamefont{Del~Debbio}},
  \bibinfo{author}{\bibfnamefont{L.}~\bibnamefont{Giusti}},
  \bibinfo{author}{\bibfnamefont{M.}~\bibnamefont{L{\"u}scher}},
  \bibinfo{author}{\bibfnamefont{R.}~\bibnamefont{Petronzio}},
  \bibnamefont{and} \bibinfo{author}{\bibfnamefont{N.}~\bibnamefont{Tantalo}},
  \bibinfo{journal}{JHEP} \textbf{\bibinfo{volume}{02}}, \bibinfo{pages}{011}
  (\bibinfo{year}{2006}), \eprint{hep-lat/0512021}.

\bibitem[{\citenamefont{Farchioni
  et~al.}(2005{\natexlab{a}})}]{Farchioni:2004us}
\bibinfo{author}{\bibfnamefont{F.}~\bibnamefont{Farchioni}}
  \bibnamefont{et~al.}, \bibinfo{journal}{Eur. Phys. J.}
  \textbf{\bibinfo{volume}{C39}}, \bibinfo{pages}{421}
  (\bibinfo{year}{2005}{\natexlab{a}}), \eprint{hep-lat/0406039}.

\bibitem[{\citenamefont{Farchioni
  et~al.}(2005{\natexlab{b}})}]{Farchioni:2004fs}
\bibinfo{author}{\bibfnamefont{F.}~\bibnamefont{Farchioni}}
  \bibnamefont{et~al.}, \bibinfo{journal}{Eur. Phys. J.}
  \textbf{\bibinfo{volume}{C42}}, \bibinfo{pages}{73}
  (\bibinfo{year}{2005}{\natexlab{b}}), \eprint{hep-lat/0410031}.

\bibitem[{\citenamefont{Jansen et~al.}(2007)}]{Jansen:2007sr}
\bibinfo{author}{\bibfnamefont{K.}~\bibnamefont{Jansen}} \bibnamefont{et~al.},
  \bibinfo{journal}{PoS} \textbf{\bibinfo{volume}{LAT2007}},
  \bibinfo{pages}{036} (\bibinfo{year}{2007}), \eprint{arXiv:0709.4434
  [hep-lat]}.

\bibitem[{\citenamefont{Kamleh et~al.}(2008)\citenamefont{Kamleh, Lasscock,
  Leinweber, and Williams}}]{Kamleh:2007bd}
\bibinfo{author}{\bibfnamefont{W.}~\bibnamefont{Kamleh}},
  \bibinfo{author}{\bibfnamefont{B.}~\bibnamefont{Lasscock}},
  \bibinfo{author}{\bibfnamefont{D.~B.} \bibnamefont{Leinweber}},
  \bibnamefont{and} \bibinfo{author}{\bibfnamefont{A.~G.}
  \bibnamefont{Williams}}, \bibinfo{journal}{Phys. Rev.}
  \textbf{\bibinfo{volume}{D77}}, \bibinfo{pages}{014507}
  (\bibinfo{year}{2008}), \eprint{arXiv:0709.1531 [hep-lat]}.

\bibitem[{\citenamefont{Mignani, Rosa}(1995)}]{Mignani:1995}
\bibinfo{author}{\bibnamefont{S.}~\bibnamefont{Mignani}},
\bibinfo{author}{\bibnamefont{R.}~\bibnamefont{Rosa}},
\bibinfo{journal}{Computer Physics Communications}
\textbf{\bibinfo{volume}{92}}, \bibinfo{pages}{203-213}
(\bibinfo{year}{1995})

\end{thebibliography}


\end{document}